# PLASMAS AND GASES

# STUDY OF MULTICOMPONENT PLASMA PARAMETERS IN THE PULSED REFLEX DISCHARGE

**YU.V. KOVTUN, A.I. SKIBENKO, E.I. SKIBENKO, YU.V. LARIN, A.N. SHAPOVAL, E.D. VOLKOV, V.B. YUFEROV**

**National Science Center "Kharkov Institute of Physics and Technology", Nat. Acad. of Sci. of Ukraine**

*(1, Akademichna Str., Kahrkiv 61108, Ukraine; e-mail: Ykovtun@kipt.kharkov.ua)*

Parameters of a dense ($10^{13} - 10^{14}$ cm$^{-3}$) multicomponent gas-metal plasma in the pulsed reflex discharge with a moderate power ($W \leq 10$ MW) have been studied. The dynamics of the plasma density in time, the mass-element composition of the plasma generated, the radial distribution of the electron density in plasma, the rotation velocity and the rotation frequency of a plasma layer with $n_p \geq n_{crit}$, the radial electric field strength, and the recombination factor at the stage of plasma density decay in the discharge have been determined. The plasma particle separation factor has been evaluated.

## 1. Introduction

As a rule, when carrying out researches (theoretical or experimental) in plasma physics, the simplest model of two-component completely ionized plasma which consists of electrons and identical ions of charge $Z$ is selected [1, 2]. However, real plasma, which is used in a number of experimental and technical installations, as well as natural plasma (for instance, ionospheric one), is practically always multicomponent. For example, even in rather "pure" hydrogen plasma generated in installations for thermonuclear fusion, an appreciable role belongs to impurities that arrive from the discharge chamber walls and considerably influence the magnitudes of the radiation emission, electroconductivity, and other plasma parameters [3]. The multicomponent composition of plasma must be taken into account, when studying plasma in MHD-generator channels, thermoemission converters, magnetoplasma separators, and in many other cases. The multicomponent composition of plasma is connected with the presence of several kinds of ions or neutral particles in it. The course and, hence, the description of many processes—e.g., mass and energy transfer—become much more complicated in multicomponent plasma systems, which can result in a considerable complication of the analysis of such systems [1,2]. Therefore, it is interesting and challenging to organize and carry out experimental researches dealing with the properties of multicomponent controllable plasma. Moreover, the interest in such researches is enhanced by practical needs in obtaining this information for developing a number of magnetoplasma technologies, including the author's ones [4–7]. For instance, in work [5], a device was proposed to separate a substance into its elements, which was based on the reflex discharge. The characteristic features of this proposition are as follows. (1) Cathodes either consist of a substance to be separated or contain it. (2) The process of plasma formation includes two stages. At the first stage, preliminary plasma is generated by means of the ionization of an ignition gas. At the second stage, the separated substance generates the main (working) plasma. (3) Streams and particles of plasma, which was preliminarily formed from the ignition gas, bombard the external surfaces of cathodes, thus supplying the discharge with sputtered particles of the cathode material (the separated substance). In essence, the creation concept for such a device was formulated in work [5]. Its partial experimental verification was carried out in this work. Thus, this work aimed at studying the parameters of dense ($10^{13}$–$10^{14}$ cm$^{-3}$) multicomponent plasma in the pulse reflex discharge with a moderate power of 1 to 10 MW. In particular, these are the element composition of plasma, plasma density dynamics at every stage of discharge evolution (growth, existence, and decay), radial distributions of the den-

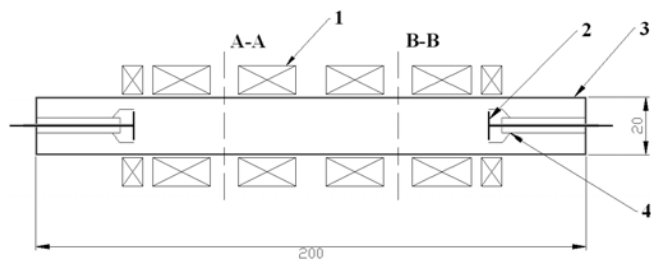

Fig. 1. Scheme of the installation for studying the pulse reflex discharge: *1* – magnetic system, *2* – cathode, *3* – vacuum chamber, *4* – insulator; A–A and B–B are cross-sections, where diagnostic ports are located

sity, rotation velocity, rotation frequency, and some others.

## 2. Experimental Installation

The parameters of plasma in the pulse reflex discharge were studied on an installation presented in Fig. 1. The discharge chamber had the following dimensions: the internal diameter of 20 cm and the length of 200 cm.

The magnetic field was created by a solenoid consisting of six coils. Two of them, the end ones, created magnetic mirrors with a mirror ratio of 1.25, and four medium coils generated a magnetic field homogeneous along the chamber axis, with a maximal strength $H_0 \leq 9$ kOe at the supply voltage $U_H \leq 3$ kV obtained from the capacitor bank. The duration of a magnetic field pulse was 18 ms. The initial pressure in the chamber was $10^{-6}$ Torr; then the ignition gas (argon or the gas mixture of krypton (88.89%), xenon (7%), nitrogen (4%), and oxygen (0.1%)) was inlet until a pressure of $(1–8) \times 10^{-3}$ Torr was reached. Gas-metal plasma was generated as a result of the discharge in the environment consisting of the ignition gas and the sputtered cathode material. The diameter of cathodes was 100 mm. They were fabricated from a composite material, namely, copper with a vacuum-arc-deposited titanium coating [8]. The conditions and regime of the cathode sputtering were as follows: Cu was the substrate material; high-purity Ti was a sputtered substance; Ar at $p = 1 \times 10^{-5}$ Torr was the sputtering environment; the sputtering time was up to 30 min; sputtering was carried out on a Bulat-6 serial installation; and the thickness of a polycrystalline coating varied from 2–2.5 to 5 $\mu$m. The substrate surface was preliminary cleaned in the environment of UHF plasma at the Ar pressure $p = 2 \times 10^{-2}$ Torr and a bias voltage of 1000 V.

To study the parameters of gas-metal plasma in the pulse reflex discharge with $n_p = 10^{12} \div 10^{14}$ cm$^{-3}$, a number of diagnostic techniques were used: (1) UHF interferometry at the wavelengths $\lambda = 8$ and 4 mm, (2) UHF reflectometry at the same wavelengths and provided $n_p \geq n_{\rm crit}$, (3) correlation reflectometry for the determination of the plasma rotation velocity [9], (4) optical spectrometry in the wavelength range $\lambda = 220 - 680$ nm, and (5) a Rogowski belt. The indicated diagnostic means were placed in diagnostic ports, except for position (5), which were located in two transverse cross-sections of the vacuum chamber, as is illustrated in Fig. 1. They were used to measure the following parameters and characteristics of the plasma formed in the discharge: (1) an UHF interferometer was used to measure the time dependence of the average plasma density in the range $10^{12} - 7 \times 10^{13}$ cm$^{-3}$; (2) the electron density profile was determined from measurements of the wave reflection at one frequency and the average density variations at another frequency, as well as by measuring the average density at various chords [10]; (3) the velocities of plasma poloidal rotation were determined by registering the fluctuations of signals reflected from the layers with the same density, which were spaced in the poloidal direction, and the autocorrelation or cross-correlation function of those signals was studied; (4) the electric field strength was determined from the measured velocity of plasma rotation taking the radial distribution of the magnetic field strength into account, (5) the element composition of the plasma formed was measured with the use of the spectrometry method, (6) the discharge current was determined with the help of a Rogowski belt.

## 3. Experimental Results and Their Discussion

In this work, to determine the element composition of the generated plasma, we measured the line spectrum of the plasma radiation in the wavelength range $\lambda = 220 \div 680$ nm. The integrated, within the discharge time, spectrum of the argon-titanium plasma radiation is depicted in Fig. 2.

The major plasma components are argon (the working gas), nitrogen, oxygen, and OH (the remnants of a residual atmosphere in the discharge chamber), as well as C, Fe, Cr, and Ti. A probable source of carbon atoms in the discharge can be the oil vapor which migrated from the diffusion pump and became deposited on the vacuum chamber walls. High-molecular hydrocarbonic compounds get into the discharge from the walls in the desorption and sputtering regime. Afterward, the high-molecular compounds dissociate in a discharge and become ionized, which results in the forma-tion of atomic carbon. The presence of Fe and Cr testifies to the sputtering of internal constructional elements in the discharge chamber. Titanium in the spectrogram corresponds to the material of the sputtered cathode layer, which penetrates into the discharge in a significant amount (to 50% and more) under the influence of the corpuscular sputtering, cathode spots (arches), and blistering.

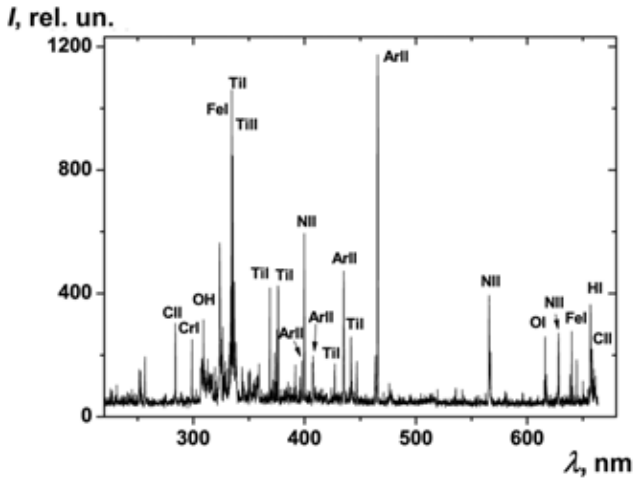

Fig. 2. Spectrogram of Ar + Ti plasma ($U_{dis} = 3.5$ kV, $U_H = 1.5$ kV, $p = 2 \times 10^{-3}$ Torr)

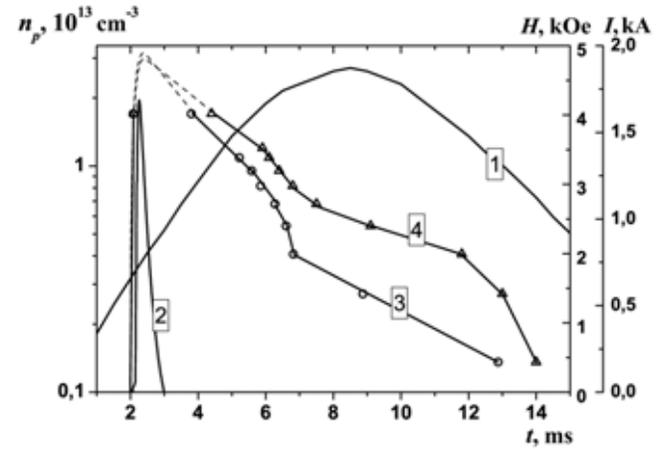

Fig. 3. Time dependences of magnetic field (*1*), discharge current (*2*), and average plasma density for Ar + Ti (*3*,○) and Kr − Xe − N$_2$ − O$_2$ + Ti (*4*,△) mixtures. $U_{dis} = 3.5$ kV, $U_H = 1.5$ kV, $p = 2 \times 10^{-3}$ Torr

In Fig. 3, the time dependences of the average plasma density in the mixtures Ar+Ti and Kr−Xe−N$_2$−O$_2$+Ti under identical initial conditions are shown. The discharge at a voltage of 3.5 kV across the working gas (mixture) was initiated with a 2-ms delay after the magnetic field had been switched-on (Fig. 3, curve *1*). The duration of a discharge current pulse and its maximal value, $I_{dis}$, were about 1 ms and 1.7 kA, respectively (Fig. 3, curve *2*). Measurements at the wavelength $\lambda = 4$ mm demonstrated that the average plasma density is more than or equal to $7 \times 10^{13}$ cm$^{-3}$, which evidences a high (close to 100%) level of ionization in the working substance. Conditionally, the time evolution of the average gas-metal plasma density can be divided into three stages. The first stage corresponds to the generation of plasma and the growth of its density to a value of about $1.7 \times 10^{13}$ cm$^{-3}$. This stage lasts for the following time intervals: about 90 ms for the mixture Ar+Ti and about 78 ms for the mixture Kr − Xe − N$_2$ − O$_2$ + Ti. The second stage corresponds to the existence of plasma with the densities $n_p \geq 1.7 \times 10^{13}$ cm$^{-3}$. Here, the characteristic time intervals are $1.2 - 1.8$ ms (Ar + Ti) and $2 - 2.5$ ms (Kr − Xe − N$_2$ − O$_2$ + Ti). It is worth noting that, at this stage, the plasma density can reach values more than $10^{14}$ cm$^{-3}$. The next third stage corresponds to the reduction and the decay of the plasma density within the $5 - 12$-ms interval.

To describe the processes that occur in the reflex discharge at the initial stage of gas ionization, the following set of equations can be used:

$$n_p = n_e v_e \sigma_e n_0 t, \tag{1}$$

$$\frac{dn_p}{dt} = \langle v_e \sigma_e \rangle n_0 n_p. \tag{2}$$

They were used earlier to calculate and analyze the initial stage of plasma formation in the beam-plasma discharge [11] and in the electric discharge in a fluid [12]. The notations used are: $n_p$ is the current plasma density dependent on the time $t$; $n_e$ and $v_e$ are the density and the average velocity, respectively, of the electron flow; $n_0$ is the density of a neutral medium; and $\sigma_e$ is the ionization cross-section of neutral gas molecules by electrons [13–15]. Equation (1) describes the linear stage of plasma formation stimulated by the shock ionization of a gas-metal mixture by particles of the initial electron flow. Equation (2) deals with the exponential stage of the discharge, when the ionization of a neutral medium occurs, mainly, due to plasma electrons accelerated owing to their collective interaction [16]. For the sake of comparison, in Figs. 4,*a* and *b*, the experimental (curves *1*) and theoretical (curves *2* and *3*) variations of the plasma density in time are depicted. Replacing the quantity $t$ in Eq. (1) by the experimental plasma lifetime $\tau$ (i.e. a time interval, after which the plasma density be-comes $e$ times lower), we obtain the value for the equilibrium density of plasma created by the electron flow. The results of calculations for a Kr gas and the Kr–Xe–N$_2$–O$_2$ gas mixture showed that the difference between the periods needed for the density to achieve identical values in both cases is at a level of 0.01%; therefore, Fig. 4,b illustrates the calculation results for pure Kr only. In the cases of argon and krypton, the addition of titanium to obtain its content of 50% (Figs. 4,a and b; curves 3) results in an insignificant increase of the density growth time.

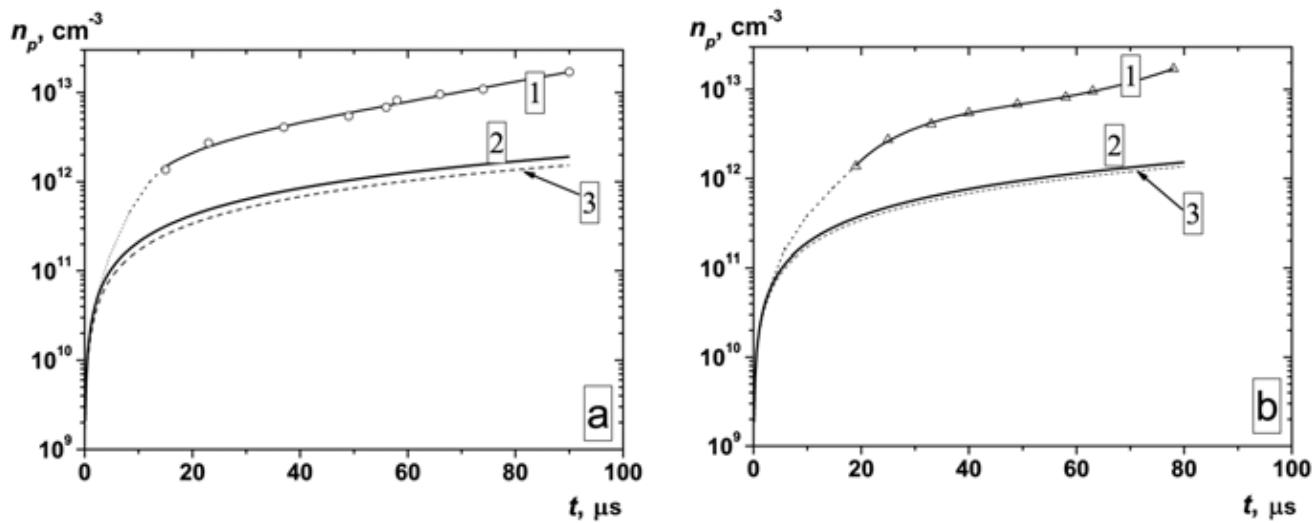

Fig. 4. Comparison of experimental and theoretical results for the time evolution of the plasma density in various gas-metal mixtures: (a) Ar + Ti and (b) Kr − Xe − N$_2$ − O$_2$ + Ti. 1 – experiment ($U_{\mathrm{dis}} = 3.5$ kV, $U_H = 1.5$ kV, $p = 2 \times 10^{-3}$ Torr), 2 – calculation results for a gas target at an electron energy of 3.5 keV, 3 – the same as for curve 2, but with a titanium content of 50%

However, the generated plasma density exceeds the density in the initial electron flow by more than an order of magnitude. Therefore, the further ionization of molecules of the neutral medium will be carried out by secondary electrons of plasma. In this case, the density evolution is described by Eq. (2). In so doing, it is assumed that the time of electron acceleration in plasma is much less than the plasma density growth time, and plasma losses at $n_p \ll n_0$ are negligibly low. Such an approximation is valid only for times $t \ll \tau$. Later on, the plasma density growth becomes more and more driven by a balance between the density growth and plasma losses, in accordance with Eq. (2).

Among the processes that result in plasma losses, the recombination, plasma diffusion across the magnetic field, and plasma leakage through the mirror dominate. The extrapolation of experimental curve 1, $n_p = f(t)$, into the region of low plasma densities up to its intersections with analogous theoretical curves 2 and 3 gives values for the threshold plasma density, after reaching which the plasma density starts to grow exponentially. Those values are equal to about $4 \times 10^{10}$ cm$^{-3}$ for both pure-gas (e.g., Ar or the mixture Kr − Xe − N$_2$ − O$_2$) and gas-metal (Ar + Ti and Kr − Xe − N$_2$ − O$_2$ + Ti) targets (media). The time of reaching the threshold density was $2.5 \times 10^{-6}$ s for Ar-based working substances and $2 \times 10^{-6}$ s for Kr − Xe − N$_2$ − O$_2$.

The dependences of the lifetime $\tau_{\mathrm{cutoff}}$ of multicomponent plasma with $n_p \geq 1.7 \times 10^{13}$ cm$^{-3}$ on the initial pressure of the ignition gas are depicted in Fig. 5. During the time $\tau_{\mathrm{cutoff}}$, the plasma density reaches its maximal value. Afterward, particles escape from the adiabatic trap owing to their loss through the loss cone, diffusion across the magnetic field, and electron–ion recombination, so that the plasma density starts to decrease. Recharge processes result in a reduction of the plasma energy content and the radiation cooling, as well as in a modification of the element and charge compositions of plasma ions.

We studied the spatial plasma characteristics of the mixtures Ar + Ti and Kr − Xe − N$_2$ − O$_2$ + Ti. The spatial distribution of the plasma density was determined by measuring the maximal density found from the probe frequency reflection, and the average density was measured by a phase incursion of the probe frequency. This method can be realized in two versions [17, 18]: 1) by using the total phase incursion of a wave induced by the density variation from 0 to the value corresponding to the reflection which occurs when the maximal criti-

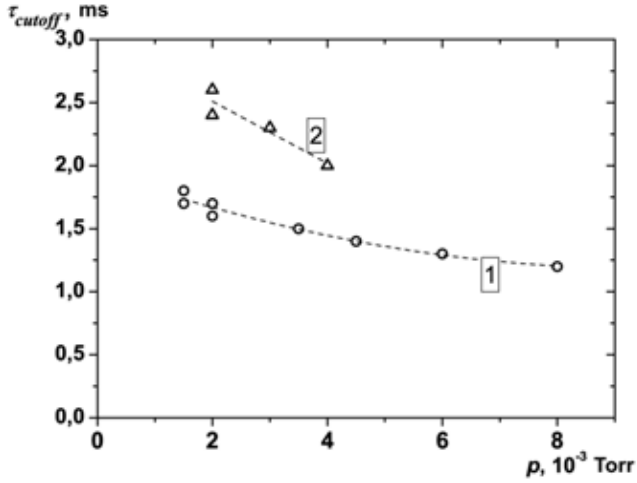

Fig. 5. Dependences of the existence time of plasma with the density $n_p \geq 1.7 \times 10^{13}$ cm$^{-3}$ on the working gas pressure ($U_{dis} = 3.5$ kV, $U_H = 1.5$ kV): Ar + Ti (*1*, ○) and Kr − Xe − N$_2$ − O$_2$ + Ti (*2*, △)

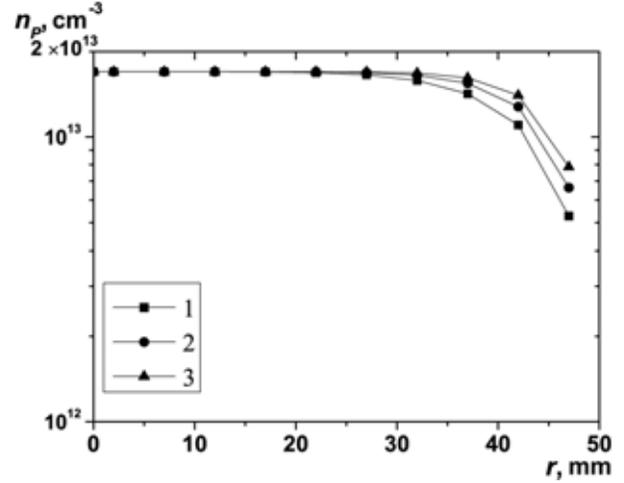

Fig. 6. Radial distribution of the electron density in plasma ($U_H = 1.5$ kV, $p = 2 \times 10^{-3}$ Torr). Mixture Ar + Ti. (*1*) $U_{dis} = 3.2$ kV ($\gamma = 6$), (*2*) $U_{dis} = 3.5$ kV ($\gamma = 8$), (*3*) $U_{dis} = 3.8$ kV ($\gamma = 10$)

cal density in a profile is attained, and 2) by finding the reflection radius for a certain wave ($n_p = n_{crit}$) and by simultaneously measuring the average density at a higher frequency using the interference of an UHF signal. In this case, we assumed the density distribution over the device radius to be a power-law function

$$n_p(r) = n_{max}\left[1 - \left(\frac{r}{r_{max}}\right)^{\gamma}\right], \qquad (3)$$

where $r_{max}$ is the maximal radius of a plasma aggregate, and $\gamma$ is the power exponent. The distance $r$ between the plasma boundary and the plasma layer with the critical density was determined from the reflected signal phase:

$$\Phi = \frac{2\omega}{c}\int\limits_0^{r_{crit}}\sqrt{1 - \frac{n(r)}{n_{crit}}}dr - \frac{\pi}{2}, \qquad (4)$$

where $n_{crit}$ is the critical density for a probe wave, $n(r)$ the plasma density at the point $r$, $\omega$ the probe wave frequency, and $c$ the velocity of light. By analyzing the experimental results, the maximal radius of a plasma layer with the critical density $n_{crit} \geq 1.7 \times 10^{13}$ cm$^{-3}$ was found to be equal to 3.6–4.2 cm. The rate of plasma layer radius growth to its maximal value was about $(5 \div 15) \times 10^3$ cm/s. The radial density profile was determined in the form of Eq. (3) in a vicinity of the cutoff wavelength $\lambda = 8$ mm, at which $n_p = n_{crit}$. This quantity can be less than $n_{crit}$ by 10–20% owing to the probe flow refraction. Taking this fact and Eq. (3) into consideration, we obtain that the phase shift of a probe wave, when the plasma

density varies within the interval from 0 to $n_{crit}$, is

$$\frac{\Delta\Phi}{2\pi} = \frac{2r_{max}}{\lambda} \times$$

$$\times\sqrt{1 - \int\limits_0^1\left[1 - \frac{n_{max}}{n_p}\left[1 - \left(\frac{r}{r_{max}}\right)^{\gamma}\right]\right]d\frac{r}{r_{max}}} =$$

$$= \frac{2r_{max}}{\lambda}\frac{\gamma}{\gamma + 2}. \qquad (5)$$

One can see that $(\Delta\Phi/2\pi) \rightarrow (2r_{max}/\lambda)$ as $\gamma$ grows, i.e. when the profile approaches the uniform one. In Fig. 6, the profiles obtained by the method described above are presented. They correspond to the moments $t$, when the wave with $\lambda = 8$ mm started to pass at different discharge voltages $U_{dis}$, so that those moments were also different. With increase in $U_{dis}$, the power exponent $\gamma$ changed in the range from 6 to 10. Therefore, the profiles obtained experimentally for the spatial distribution of the plasma density are acceptable from the viewpoint of achieving the maximal separator productivity. Similar results for a monogas-based plasma were obtained in work [19].

With help of correlation reflectometry, we determined the rotation velocity of plasma in the layer with the density $n_p \geq n_{crit}$ [9, 20]. In the general case, the velocity of plasma poloidal rotation can be found from the relation

$$v = \frac{\Delta l}{\Delta t} = \frac{\left(\frac{\Delta\varphi(r_1 + r_2)}{2}\right)}{\Delta t}, \qquad (6)$$

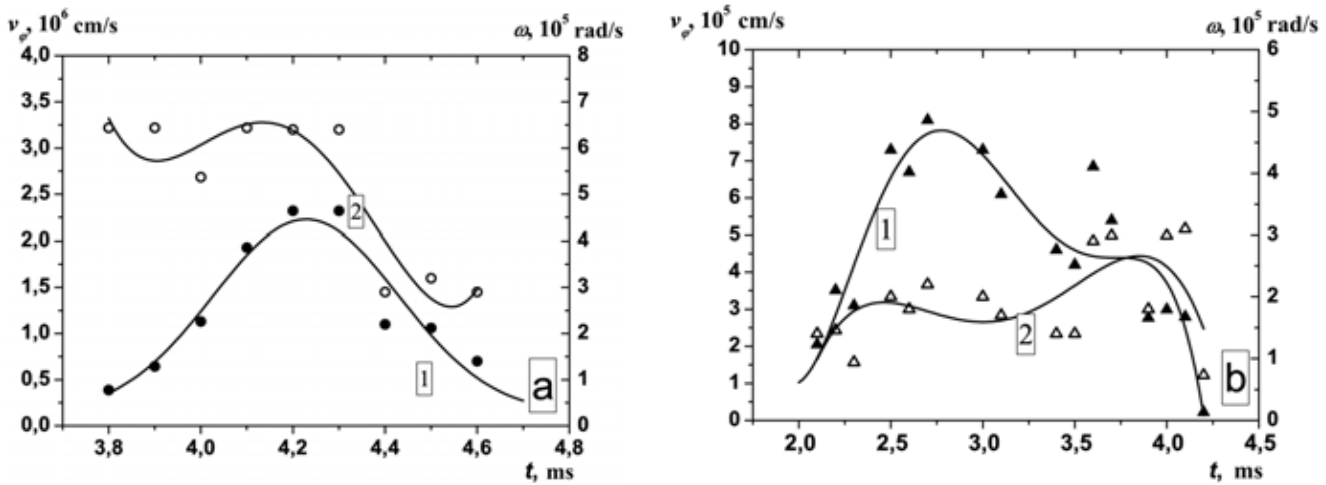

Fig. 7. Time dependences of the rotation velocity ($1$, $\bullet$, $\blacktriangle$) and the frequency ($2$, $\circ$, $\triangle$) of a plasma layer with $n_p \geq 1.7 \times 10^{13}$ cm$^{-3}$. ($a$) Ar + Ti ($\bullet$, $\circ$), ($b$) Kr − Xe − N$_2$ − O$_2$ + Ti ($\blacktriangle$, $\triangle$)

where $\Delta\varphi$ is the angular distance between the receiving points of a reflected wave; $r_1$ and $r_2$ are the layer positions determined from the phase shift of the reflected wave (in the case of a profile with circular symmetry, $r_1 = r_2$); $\Delta t$ is either a time shift of the cross-correlation function (CCF) maximum or the period of the autocorrelation function (ACF). The period of ACF corresponds to the duration of one turn by the plasma layer. The difference between the ACF and CCF regimes consists in that the determination of a plasma layer rotation direction is possible only in the CCF regime.

The plasma is ranged with the use of an ordinary wave with the same frequency; this wave is reflected from a layer with identical electron density. The method does not allow one to distinguish between the rotation velocities of individual plasma components with different mass numbers. Using the CCF and ACF, as well as the layer radius, we determined the velocity and the frequency of plasma rotation in crossed $E$ and $H$ fields (Fig. 7). The direction of plasma layer rotation, which was determined on the electron level from a displacement of the maximum of the cross-correlation function, did not coincide with the direction of ion cyclotron rotation. The maximal rotation velocity was $2.3 \times 10^6$ cm/s for the Ar + Ti mixture and $8.1 \times 10^5$ cm/s for the Kr − Xe − N$_2$ − O$_2$ + Ti one. Under experimental conditions, the Larmor radius of Ti ions with an energy of $1 - 10$ eV did not exceed the measured radius of the plasma layer with $n_p \geq 1.7 \times 10^{13}$ cm$^{-3}$. It is worth noting that the velocity at the periphery of a plasma filament, as well as near the cathodes, is confined by the critical value [21] originated from the interaction between plasma and neutral atoms. The values of the critical velocity can be determined from the relation $v_c = (2e\varphi_i/m_i)^{1/2}$, where $\varphi_i$ is the ionization potential. For Ar and Kr, this quantity equals $8.7 \times 10^5$ and $5.6 \times 10^5$ cm/s, respectively. The critical velocity can be exceeded, provided that the ionization is complete, which was observed in experiment.

Under the influence of crossed electric and magnetic fields, particles undergo the action of various forces. In particular, in the radial direction, i.e. a direction perpendicular to the particle rotation axis, these forces are the centrifugal force $F_c$ caused by the particle motion, the electric force $F_E$ due to the electric field $E_r$ acting on a particle, and the magnetic force $F_B$ due to the magnetic field $B_z$. Those forces can be expressed as follows: $F_c = mv_\varphi^2/r$, $F_E = eE_r$, and $F_B = ev_\varphi B_z$, where $m$ is the particle mass, $r$ the transverse dimension of the plasma layer, $v_\varphi$ the rotation velocity, $E_r$ the strength of the radial electric field, and $B_z$ the magnetic field induction. In the reflex discharge (the Penning discharge), the electric field is directed radially inward, which means that the positive voltage increases as the distance from the axis grows. Under such conditions, the electric force opposes the action of the centrifugal force. In a stationary state, the velocities of electrons and ions are determined from the equilibrium condition

$$\frac{mv_\varphi^2}{r} - eE_r - ev_\varphi B_z = 0. \qquad (7)$$

Whence it follows that the rotation velocity of the electron plasma component is $v_\varphi = -E_r B_z$. For ions, taking the centrifugal force into account [22], we have

$$v_\varphi = \frac{r\omega_{ci}}{2}\left(1 + \sqrt{1 + \frac{4eE_r}{\omega_{ci}rB_z}}\right), \qquad (8)$$

where $\omega_{ci}$ is the cyclotron frequency. Those formulas can be used to evaluate the strength of the radial electric field in plasma. For the studied mixtures, the values obtained fall within the interval 22–140 V/cm for $Ar + Ti$ and 1–30 V/cm for $Kr - Xe - N_2 - O_2 + Ti$, the specific value depending on discharge conditions.

In the rotating plasma, the centrifugal effects may bring about a spatial separation of ions. Moving along the azimuth, the particles undergo the influence of the centrifugal force; as a result, the centrifugal drift of particles arises. Under the action of the centrifugal force, ions with different masses obtain different velocities along the azimuth. This gives rise to the situation where light ions move to the center, whereas heavy ions become shifted toward the periphery. Both fluxes are equal to each other. The ion separation occurs until the centrifugal force acting on a unit volume is compensated by the pressure gradient. The separation coefficient $a$ is defined as the ratio between the numbers of particles of the substance, which is to be separated, at the periphery and at the center of a plasma column. In accordance with work [23], this parameter is calculated by the formula

$$a = \exp\left(\frac{\Delta m v_\varphi^2}{2kT}\right), \qquad (9)$$

where $\Delta m$ is the mass difference between separated elements (isotopes), $v_\varphi$ the rotation velocity, $T$ the plasma temperature, and $k$ the Boltzmann constant. The plasma temperature was determined earlier, and the following results for its components were obtained: $T_e \leq 10 - 20$ eV and $T_i \leq 1 - 2$ eV at $n_p \approx 10^{14}$ cm$^{-3}$. Using the experimentally obtained velocity of plasma rotation (Fig. 7) and adopting $T_i \approx 10$ eV for the estimation of a minimal value of the plasma particle separation coefficient $a$, we obtain $a \approx 9$ for the mixture $Ar + Ti$ and $a \approx 4$ for the mixture $Kr + Ti$.

At the stage of discharge decay, when having determined the time dependence of the average plasma density and the density distribution over the radius, one can find the recombination coefficient $a_r$. The equation describing the plasma density evolution in time can be written down in the form

$$\frac{dn(r,t)}{dt} = D\Delta n(r,t) - \alpha_r n^2(r,t), \qquad (10)$$

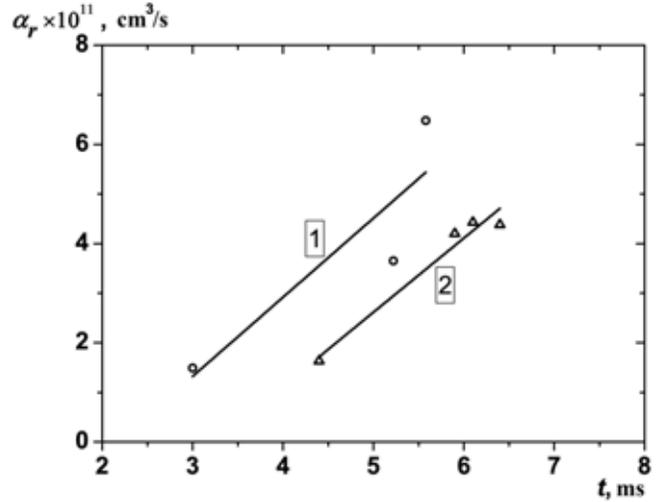

Fig. 8. Time evolution of the recombination coefficient: (1) $Ar + Ti$ ($\circ$), (2) $Kr - Xe - N_2 - O_2 + Ti$ ($\triangle$)

where $D\Delta n(r,t)$ and $\alpha_r n^2(r,t)$ are diffusion and recombination losses, respectively.

In the absence of the diffusion, a reduction of the plasma density in time owing to the recombination is described as follows:

$$n^e(r,t) = \frac{n_0^e(r)}{\alpha_r t n_0^e(r) + 1}, \qquad (11)$$

where $n_0^e(r)$ is the density distribution at the initial time moment (for instance, it can be the moment, when the probe signal starts to pass). The same relation is also valid for the plasma density averaged over the radius:

$$n^e(t) = \frac{n_0^e}{\alpha_r t n_0^e + 1}. \qquad (12)$$

The time dependence of the inverse plasma density $1/n_p$ (see Fig. 3) testifies that this quantity varies linearly at the initial stage of the plasma decay (within the time interval from 4 to 5.5 ms for $Ar + Ti$ and from 4 to 7.5 ms for $Kr - Xe - N_2 - O_2 + Ti$). Within the time interval 5.5 − 13 ms for $Ar + Ti$ and 7.5 − 14 ms for $Kr - Xe - N_2 - O_2 + Ti$, the quantity $1/n_p$ varies exponentially. According to work [24], this fact testifies that the recombination losses dominate during the first time interval, whereas another mechanism—diffusion losses—starts to prevail at the second stage. The time dependence of the recombination coefficient calculated by formula (1) is depicted in Fig. 8.

## 4. Conclusions

The researches of the parameters of multicomponent dense ($n_p = 10^{13} \div 10^{14}$ cm$^{-3}$) plasma in the pulse reflex discharge with a moderate power ($U_{dis} = 3.5$ kV, $I_{dis} = 1.7$ kA, $W \approx 6$ MW, $Q = 6$ kJ) at a power contribution per plasma unit volume lower than or equal to 0.58 J/cm$^3$ have been carried out. The following parameters and characteristics of multicomponent plasma formed in the discharge have been determined.

1. The time dependence of the multicomponent plasma density at every stage of discharge evolution (growth, existence, decay).

2. The mass-element composition of the generated plasma which contains ions and atoms of Ar, Ti, N, O, C, Fe, Cr, and OH.

3. From the reflection of an UHF wave and a variation of the average density of the electron plasma component determined at another frequency (wave), as well as from the average density variations along different chords, the profile of a radial distribution of the electron density in plasma has been determined. It turned out that the electron density distribution in plasma over the radius corresponds to the power-law function (3), the power exponent $\gamma$ of which varies from 6 to 10, depending on the discharge parameters.

4. With the help of a reflectometer, by ranging the plasma with an ordinary wave at the same frequency and by analyzing the reflection from a layer with identical electron density, the velocity and the frequency of rotation of the plasma layer with $n_p \geq n_{crit}$ have been determined. Using this information, the radial electric field strength in plasma has been estimated.

5. Taking the influence of the centrifugal force into account, the separation coefficient $a$ of plasma particles has been estimated for the Ar − Ti ($a \approx 9$) and Kr − Ti ($a \approx 4$) mixtures.

6. The recombination coefficient $\alpha_r \approx (1.5 \div 6.5) \times 10^{-11}$cm$^3$/s has been estimated at the initial stage of the plasma density decay in the discharge.

ДОСЛІДЖЕННЯ ПАРАМЕТРІВ
БАГАТОКОМПОНЕНТНОЇ ПЛАЗМИ
В ІМПУЛЬСНОМУ ВІДБИВНОМУ РОЗРЯДІ

*Ю.В. Ковтун, А.І. Скибенко, Є.І. Скібенко, Ю.В. Ларін,*
*А.М. Шаповал, Є.Д. Волков, В.В. Юферов*

Р е з ю м е

Проведено дослідження параметрів густої $(10^{13}\text{–}10^{14}$ см$^{-3})$ газометалевої багатокомпонентної плазми імпуль-сного відбивного розряду помірної потужності ($W \leq$ 10 МВт). Визначено динаміку густини плазми в часі, масово-елементний склад утворюваної плазми, профіль радіального розподілу електронної густини плазми, швидкість та частоту обертання плазмового прошарку з $n_p \geq n_{\text{crit}}$, величину напруженості радіального електричного поля, коефіцієнт рекомбінації на стадії розпаду густини плазми в розряді, оцінено величину коефіцієнта розділення частинок плазми.